\documentstyle[12pt]{article}

\begin{document}

\begin{flushright}
IMSc/98/12/56 \\
hep-th/9812043
\end{flushright} 

\vspace{2mm}

\vspace{2ex}

\begin{center}
{\large \bf  

             Holographic Principle during Inflation and \\

\vspace{2ex}
             a Lower Bound on Density Fluctuations          } \\ 

\vspace{8ex}

{\large  S. Kalyana Rama, \footnote{On leave of absence from 
Mehta Research Institute, Chhatnag Road, Jhusi, 
Allahabad 221 506, India.} Tapobrata Sarkar}

\vspace{3ex}

Institute of Mathematical Sciences, C. I. T. Campus, 

Taramani, CHENNAI 600 113, India. 

\vspace{1ex}

email: krama@imsc.ernet.in , sarkar@imsc.ernet.in\\ 
\end{center}

\vspace{4ex}

\begin{quote}
ABSTRACT.  
We apply the holographic principle during the inflationary stage
of our universe. Where necessary, we illustrate the analysis in
the case of new and extended inflation which, together, typify
generic models of inflation. We find that in the models of
extended inflation type, and perhaps of new inflation type also,
the holographic principle leads to a lower bound on the density
fluctuations.
\end{quote}

\vspace{2ex}


\newpage

\vspace{4ex}

{\bf 1.} 
The holographic principle is simple and yet profound
\cite{thooft}: it implies that the degrees of freedom in a
spatial region can all be encoded on its boundary, with a
density not exceeding one degree of freedom per planck
area. Accordingly, the entropy in a spatial region does not
exceed its boundary area in planck units. Also, for example, the
physics of the bulk is describable by the physics on the
boundary. This has, indeed, been shown recently for some anti de
Sitter spaces \cite{w,sw}.

Fischler and Susskind have proposed how to apply the holographic
principle in cosmology, and showed that our universe has evolved
in the past in accordance with this principle and will continue
to do so if it is flat or open \cite{fs}. This principle has
recently been applied in the context of pre big bang scenario in
\cite{rey}, and in the context of singularity problem in
\cite{riotto}.

Our universe is believed to have gone through an inflationary
stage in the past \cite{oi}. Among other things, an
enormous amount of entropy is released into the universe at the
end of inflation. Is this amount of entropy consistent with the
holographic principle?

We apply the holographic principle during the inflationary stage
and study this issue. The relevent details are often model
dependent. Hence, where necessary, we illustrate the analysis in
the case of new \cite{ni} and extended \cite{ei} inflation where
the reheating and entropy production is due to inflaton decay
and bubble wall collisions repectively. These two models,
together, typify generic models of inflation. For more details,
see \cite{ni,book,review1} for new and \cite{ei,ew,kst,review2}
for extended inflation.

We find that in the models of extended inflation type, and
perhaps of new inflation type also, the holographic principle
leads to an upper bound on the inflation factor. This, in turn,
leads to a lower bound on the density fluctuations in the
universe, which seed the large scale structure formation. These
consequences, although obtained explicitly for new and extended
inflation, are expected to be valid generally. Considering the
approximations involved, the lower bound on density fluctuations
we find is remarkably close to the observed value $\simeq 10^{-
6}$ \cite{cobe} if inflation takes place when the temperature of
the universe $\simeq 10^{14} GeV$. To our knowledge, this is the
first instance where a {\em lower bound} on the density
fluctuations is obtained theoretically. Such a bound, if
established rigorously and in a model independent way, could be
taken as a prediction of the holographic principle.

In the models of new inflation type, the inflaton decay is often
modelled by that of a massive scalar field interacting with
other fields. The holographic principle is automatically
satisfied in these models with no further consequences.

\vspace{4ex}

{\bf 2.}
In the context of cosmology, the Fischler-Susskind (FS) proposal
for the application of holographic principle is as follows. Let
$\Gamma$ be a spherical spatial region of coordinate size $r$
with boundary $B$ and let $L$ be the light-like surface formed
by past light rays from $B$ towards the center of $\Gamma$. Then
according to FS proposal, the holgraphic principle implies that
the entropy passing through $L$ never exceeds the area of $B$
\cite{fs}.

Let the metric of the $3 + 1$ dimensional homogeneous isotropic
flat universe be given by the line element
\begin{equation}
d s^2 = - d t^2 + R^2 (d r^2 + r^2 (d \theta^2 
+ \sin^2 \theta d \phi^2)) \; . 
\end{equation}
Let $\rho$, $p$, and $T$ be the density, pressure, and
temperature of the universe respectively. Also, let
\[
r_H = \int_0^t \frac{d t}{R}  
\; \; \; \; {\rm and} \; \; \; \; 
d_H = r_H R 
\]
be the coordinate and the physical size of the horizon
respectively.  The (constant) comoving and the (varying)
physical entropy densities, $\sigma$ and $s$ respectively, are
then given by
\begin{equation}
\sigma = \frac{\rho + p}{T} \; R^3 \equiv s R^3 \; . 
\end{equation}
The entropy $S$ contained within $\Gamma$, and the area $A$ of
the boundary $B$ are 
\begin{eqnarray}
S & = & \frac{4 \pi}{3} s d_H^3 \label{s} \\ 
A & = & 4 \pi d_H^2 \; . \label{a} 
\end{eqnarray}
According to FS proposal, the holographic principle implies that
\begin{equation}\label{hol}
S \stackrel{<}{_\sim} A \; . 
\end{equation}
The approximate order of magnitude values of various quantities
in our universe at different epochs are tabulated below
\cite{book}. All quantities, here and in the following, are in
planck units, unless mentioned otherwise.

\vspace{2ex}

\begin{tabular}{||c||c|c|c|c||}   \hline \hline
$T$   & $T_0$                  & $5.5 eV$      
      & $10^{14} GeV$          & $T_{pl}$ 
\\ \hline \hline
$R$   & $6 \times 10^{60}$     & $2 \times 10^{56}$ 
      & $10^{34}$              & $10^{29}$         
\\ \hline
$r_H$ & $1$                    & $7 \times 10^{- 3}$ 
      & $4 \times 10^{- 25}$   & $3 \times 10^{- 30}$       
\\ \hline
$d_H$ & $6 \times 10^{60}$     & $2 \times 10^{54}$ 
      & $5 \times 10^9$        & $0.3$          
\\ \hline
$s$   & $10^{- 95}$            & $2 \times 10^{- 82}$ 
      & $9 \times 10^{- 16}$   & $2$                        
\\ \hline
$S$   & $10^{88}$              & $3 \times 10^{81}$ 
      & $5 \times 10^{14}$     & $0.3$          
\\ \hline
$A$   & $4 \times 10^{122}$    & $3 \times 10^{109}$ 
      & $3 \times 10^{20}$     & $1.4$        
\\ \hline \hline
\end{tabular}

\vspace{2ex}

\noindent $T_0 = 2.75 K$ is the present temperature of the
universe and $T_{pl} = 1.2 \times 10^{19} GeV$ is the planck
temperature. Note that the constant comoving entropy density is
given, for our choice of $r_H$(present) $= 1$, by
\[
\sigma_0 \simeq 10^{87} \; . 
\]
It is clear from the above table that the holographic principle
is obeyed in our universe from planckian time upto the
present. It is obeyed indefinitely in the future too if our
universe is flat or open \cite{fs}.

\vspace{4ex}

{\bf 3.}  
Our universe is believed to have gone through an inflationary
stage at a temperature $T_b$ (usually taken to be $10^{14} GeV
\simeq 10^{- 5}$ in planck units) \cite{book}. Generically, a
small, causally connected patch of the universe inflates from
say time $t = 0$ to $t = t_e$. The scale factor $R$ grows by a
factor of $e^N$ and the universe supercools. At the end of
inflation, the universe reheats to a temperature $T_R
\stackrel{<}{_\sim} T_b$, releasing an enormous amount of
entropy. Is the amount of entropy produced consistent with the
holographic principle?

For the region with $r_H = 1$, the natural values of the entropy
$S$ and the area $A$ at the beginning of inflation are
\begin{eqnarray*}
S_b & \simeq & \sigma_b \simeq T_b^3 t_b^3 
\simeq T_b^{- 3} \; \; (10^{15}) \\ 
A_b & \simeq & t_b^2 \simeq T_b^{- 4} \; \; (10^{20}) \; , 
\end{eqnarray*}
where the numbers in the bracket are the values if 
$T_b = 10^{- 5}$. So, $S_b < A_b$ and the holographic 
principle is obeyed at the beginning of inflation. 

At the end of inflation we get, for the region with $r_H = 1$, 
\begin{eqnarray*}
S_e & \simeq & \sigma_e \simeq e^{3 N} T_b^{- 3} \\
A_e & \simeq & e^{2 N} T_b^{- 4} 
\end{eqnarray*}
where we have taken $T_R \simeq T_b$. In order to account for
the observed entropy of the universe, we require \cite{oi,book}
\[
\sigma_e \stackrel{>}{_\sim} \sigma_0 \simeq 10^{87} \; . 
\]
Hence, 
\[
e^N \stackrel{>}{_\sim} \sigma_0^{\frac{1}{3}} T_b
\; \; (10^{24}) \; . 
\]
This is the required 60 e-folding of the inflationary scenario
\cite{oi}. For these values,
\[
\frac{S_e}{A_e} \simeq e^N T_b \; \; (10^{19}) 
\]
which clearly violates the holographic principle. The required
entropy production will not violate the holographic principle,
as applied above, only if $T_b \stackrel{<}{_\sim} 10^{- 15}
\simeq 10^4 GeV$ with $e^N T_b \stackrel{<}{_\sim} 1$. However,
such a low value is unsatisfactory for other reasons
\cite{book}.

More importantly, the above application of the holographic
principle is naive and is precisely the one Fischler and
Susskind admonished against. The spatial region $\Gamma$ and the
boundary $B$, which evolve along the light-like surface $L$ into
the present ones, are marked at the end of inflation, when 
$T = T_R \simeq T_b$, by (see the table)
\[
r_H \simeq 3 \times 10^{- 30} \; T_b^{- 1} 
\; \; (10^{- 25}) \; .  
\]
For such a region, we get 
\[
\frac{S_e}{A_e} \simeq r_H T_b e^N \simeq 10^{- 30} e^N \; . 
\]
The holographic principle is then obeyed if the inflation factor
\[
e^N \stackrel{<}{_\sim} 10^{30} \; , 
\]
which is sufficient to solve all the problems in Guth's original
proposal for inflation \cite{oi}.

\vspace{4ex}

{\bf 4.}  
Typically, however, $e^N$ is of the order of $10^{100} -
10^{300}$ in extended inflation \cite{ei} and of the order of
$10^{10^3} - 10^{10^7}$ in new inflation
\cite{ni,book}. So, the above bound is a severe constraint on
inflationary models and acheiving it is likely to be unnatural,
if possible at all.  Also, the above application of holographic
principle is in the era immediately following the entropy
production, and not when the entropy is actually being produced.

The entropy is produced at the end of inflation during the
reheating process and the universe reheats to a temperature $T_R
\stackrel{<}{_\sim} T_b$, where $T_b$ is the temperature at the
beginning of inflation. The physical entropy density $s_R$
during the entropy production can be taken, on an average, to be
\cite{book} 
\[
s_R \simeq T_R^3 \; . 
\]
The holographic principle, applied during this process to a
suitable region, to be identified below, of physical size
$\simeq d$ (hence of volume $d^3$ with boundary area $d^2$),
implies that
\begin{equation}\label{holi}
S_R \stackrel{<}{_\sim} A_R  
\; \; \; \; \longrightarrow \; \; \; \; 
T_R^3 d \stackrel{<}{_\sim} 1 \; . 
\end{equation}

The actual details of the reheating and the entropy production
are model dependent \cite{book,review1,ew}. However, the
relevent physical process falls broadly in one of the two
categories where the reheating and the entropy production are
due to (1) bubble wall collisions - typified by extended
inflation \cite{ei}, or (2) the decay of the `slow rolling'
inflaton - typified by new inflation \cite{ni}. We now identify
the size $d$ in each of these cases.

\noindent(1) The true vacuum bubbles nucleate during inflation,
expand with the speed of light, and eventually percolate the
universe, thus ending the inflation. Upon percolation, the
bubble walls collide and release the energy and entropy into the
interior of the bubbles, thereby reheating the universe to a
temperature $T_R \stackrel{<}{_\sim} T_b$.  Typically, the
reheating time is of the order of the time required for light to
cross the bubble \cite{ew}. Thus, it is natural to apply the
holographic principle to the interior of each bubble. On an
average, the time between the bubble nucleation and collision is
less than or of the order of $t_e$, the duration of inflation.
The interior of the bubble is in a true vacuum state and, thus,
its size $d \simeq t_e$. With no further condition on $T_R$, the
holographic principle implies that
\begin{equation}\label{bubble}
T_R^3 t_e \stackrel{<}{_\sim} 1 \; . 
\end{equation}

\noindent(2) The inflaton slowly rolls down to its minimum and
begins to oscillate, thus ending the inflation. The oscillating
inflaton decays into other particles, releasing the energy and
entropy into the universe and, thereby, reheating it to a
temperature $T_R \stackrel{<}{_\sim} T_b$. The entropy is
produced simultaneously and everywhere in the inflated
region. Thus, it is natural to apply the holographic principle
to any region covered by a light ray starting from a point and
travelling for a time $\simeq t_R$, where $t_R$ is the duration
of reheating. The universe is in a true vacuum state during
reheating and, thus, the size of this region $d \simeq t_R$.

The universe reheats within a few Hubble time $H_e^{- 1} \simeq
t_e$ at the end of inflation. Taking $t_R \stackrel{<}{_\sim}
t_e$, and with no further condition on $T_R$, the holographic
principle implies the relation (\ref{bubble}), same as in the
previous case.

Often, the inflaton decay is modelled by that of a massive
scalar field interacting with other fields - scalars, fermions,
photons, gravitons, etc. \cite{book,review1}. In typical models,
the reheating time $t_R \simeq \gamma_d^{- 1}$, where $\gamma_d$
is the decay rate which, for $T_b = 10^{14} GeV$, is ${\cal O}
(10^{- 6} - 10^{- 12})$ in planck units depending on the model
and the decay products. Moreover, the reheating temperature
$T_R$ is related to $\gamma_d$ by
\[
T_R \simeq \sqrt{\gamma_d} 
\]
in planck units. Note that, in these models, the reheating time
$t_R$ (the reheating temperature $T_R$) has no relation to the
Hubble time $t_e$ at the end (the temperature $T_b$ at the
beginning) of inflation. The holographic relation (\ref{holi})
then implies
\begin{equation}\label{decay}
\sqrt{\gamma_d} \stackrel{<}{_\sim} 1 \; ,  
\end{equation}
a condition well satisfied in these models.

\vspace{4ex}

{\bf 5.}
We now explore the consequences of the relation (\ref{bubble}).
The reheating temperature $T_R \stackrel{<}{_\sim} T_b$ depends
only on at what temperature the inflation sets in. The duration
of inflation $t_e$ then satisfies an upper bound given by 
(\ref{bubble}). Such an upper bound on $t_e$ can be expected,
among other things, to lead to an upper bound on the inflation
factor $e^N$. This is simply because longer the duration of
inflation, larger is the expansion factor.

An upper bound on $e^N$ can, in turn, be expected to lead to a
lower bound on the density fluctuations in the universe, which
seed the large scale structure formation. This is because,
essentially the inflation dampens the quantum fluctuations of
the fields, which reenter as density fluctuations in the later
era. Hence, larger the inflation factor, more the damping of
quantum fluctuations, and thus smaller the resulting density
fluctuations.

Although the above physical reasoning is direct and simple, the
actual calculations of $t_e$ and of the density fluctuations are
quite involved. Also, to our knowledge, there is no model
independent formula which relates the density fluctuations to
the duration or the amount of inflation. Hence, we illustrate
these consequences explicitly in the context of new and extended
inflation. However, following the above reasoning, they are
expected to be valid generally. 

\vspace{4ex}

\noindent{\bf 5 a.} 
Consider the duration of inflation $t_e$ and the expansion
factor $e^N$. (For details about various expressions used
below, see \cite{ni,book,review1} for new and
\cite{ei,ew,kst,review2} for extended inflation.)

\noindent{\bf New Inflation:} Let the inflaton potential be
\begin{equation}\label{v} 
V = V_0 - \frac{\lambda}{n} \; \phi^n
\; , \; \; \; n \ge 4 \; , 
\end{equation} 
where $V_0 \equiv \frac{3 H_b^2}{8 \pi} = M^4 \simeq T_b^4$, and
$\lambda$ is a coupling constant. Equation (\ref{bubble})
implies that the inflation factor $e^N$ is restricted by an
upper bound given by 
\[ 
N \simeq H_b t_e \stackrel{<}{_\sim}
\left( \frac{T_b}{T_R} \right)^3 T_b^{- 1} \; .  
\] 
Note that with $T_b = 10^{14} GeV$ and $T_R \simeq 0.1 T_b$,
we have $N \stackrel{<}{_\sim} 10^8$, which conforms well with
the amount of inflation occuring in these models.

The duration of inflation $t_e$ is related to the coupling
constant $\lambda$ by
\[
t_e \simeq 4 \pi^2 H_b^{\frac{8}{n} - 3} \;
\left( \frac{3}{8 \pi^2 \lambda} \right)^{\frac{2}{n}} \; . 
\]
Equation (\ref{bubble}) then implies a lower bound
\begin{equation}\label{holni}
\lambda \stackrel{>}{_\sim} T_R^{\frac{3 n}{2}} 
H_b^{4 - \frac{3 n}{2}} \; .
\end{equation}

\noindent{\bf Extended Inflation:} The model is specified by a
parameter $\omega \simeq 10 - 20 \stackrel{<}{_\sim} 25$.
Equation (\ref{bubble}) implies that the inflation factor $e^N$
is restricted by an upper bound
\[
e^N \simeq t_e^{\omega + \frac{1}{2}} 
\stackrel{<}{_\sim} T_R^{- 3 (\omega + \frac{1}{2})} \; . 
\]
Note that with $T_R \simeq T_b = 10^{14} GeV$, and with $\omega
= 10$, we have $e^N \stackrel{<}{_\sim} 10^{160}$, which
conforms well with the amount of inflation occuring in these
models.

\vspace{4ex}

\noindent{\bf 5 b.} 
Consider the density fluctuations on a scale $\lambda_0 
(\simeq 10^{60}$ for the horizon) today. Let $T_0 = 2.75 K$ 
be the present temperature of the universe.

\noindent{\bf New Inflation:} The inflaton potential is given by
(\ref{v}). The density fluctuations are then given by
\begin{equation}\label{rhoni}
\frac{\delta \rho}{\rho} \simeq A_{ni} 
H_b^{\frac{n - 4}{n - 2}} \lambda^{\frac{1}{n - 2}} 
\end{equation}
where $n \ge 4$ and 
\[
A_{ni} \simeq \frac{16}{3} \left( \frac{2}{3} \ln 
\frac{H_b \lambda_0 T_0}{T_R} \right)^{\frac{n - 1}{n - 2}} \; . 
\]
With $H_b \simeq M^2 \simeq T_b^2$, equation (\ref{holni}) then
implies a lower bound on the density fluctuations:
\begin{equation}
\frac{\delta \rho}{\rho} \stackrel{>}{_\sim} A_{ni} 
\left( \frac{T_R}{T_b} \right)^{\frac{3 n}{2 (n - 2)}} 
T_b^{\frac{n}{2 (n - 2)}} \; . 
\end{equation}
For $T_b = 10^{14} GeV$, $T_R \simeq 0.1 T_b$, and $n = 4$,
$A_{ni} \simeq {\cal O} (10^2)$ and the above bound gives
\[
\frac{\delta \rho}{\rho} \stackrel{>}{_\sim} {\cal O} (10^{- 6})
\; .  
\] 
Considering the approximations involved, the above
lower bound on the density fluctuations implied by equation
(\ref{bubble}) is remarkably close to the observed value $\simeq
10^{- 6}$ \cite{cobe} if inflation takes place at $T_b \simeq
10^{14} GeV$.

\noindent{\bf Extended Inflation:} The density fluctuations are
given by
\begin{equation}\label{rhoei}
\frac{\delta \rho}{\rho} \simeq A_{ei} 
\left( T_0 \lambda_0 \sqrt{2 \omega + 1} 
\right)^{\frac{4}{2 \omega - 1}} 
(2 \omega + 1)^{\frac{3}{2}} 
t_e^{- \frac{2 \omega + 1}{2 \omega - 1}} 
\end{equation}
where 
\[
A_{ei} \simeq \frac{\sqrt{\pi}}{3} 
\left( \frac{8 \pi}{9} \right)^{\frac{2}{2 \omega - 1}} 
\left( \frac{6 \omega + 9}{6 \omega + 5} 
\right)^{\frac{2 \omega + 3}{2 \omega - 1}} \; .  
\]
With $\lambda_0 \simeq 10^{60}$, $T_0 \simeq 10^{- 32}$, and
$\omega \simeq 10$, equation (\ref{bubble}) then implies a lower
bound on the density fluctuations:
\begin{equation}
\frac{\delta \rho}{\rho} \stackrel{>}{_\sim} 10^8 A_{ei} 
T_R^{ \frac{3 (2 \omega + 1)}{2 \omega - 1}} \; . 
\end{equation}
For $T_R \simeq T_b \simeq 10^{14} GeV$ and $\omega \simeq 10$,
$A_{ei} \simeq {\cal O} (1)$ and the above bound gives
\[
\frac{\delta \rho}{\rho} \stackrel{>}{_\sim}
{\cal O} (10^{- 7}) \; . 
\]
Considering the approximations involved, the above lower bound
on the density fluctuations implied by equation (\ref{bubble})
is remarkably close to the observed value $\simeq 10^{- 6}$
\cite{cobe} if inflation takes place at 
$T_b \simeq 10^{14} GeV$.

\vspace{4ex}

{\bf 6.}  
We conclude with a few remarks. We have shown, in the case of
new and extended inflation which, together, typify generic
models of inflation, that equation (\ref{bubble}) leads to an
upper bound on the inflation factor. This, in turn, leads to a
lower bound on the density fluctuations. As discussed before,
these consequences are expected to be valid generally.
Considering the approximations involved, the lower bound on the
density fluctuations, obtained in specific cases here, is
remarkably close to the observed value if inflation takes place
at $T \simeq 10^{14} GeV$. To our knowledge, this is the first
instance where a {\em lower bound} on the density fluctuations
is obtained theoretically. Such a bound, if established
rigorously and in a model independent way, could be taken as a
prediction of the holographic principle.

Equation (\ref{bubble}), which led to these bounds, arises as 
a consequence of the holographic principle in those models,
typified by extended inflation, where the reheating and the
entropy production are due to bubble wall collisions. The
reheating temperature is taken to depend only on at what
temperature the inflation sets in. Thus, in such models, the
lower bound on density fluctuations is a consequence of the
holographic principle.

Equation (\ref{bubble}) also arises as a consequence of the
holographic principle in those models, typified by new
inflation, where the reheating and the entropy production are
due to the decay of the `slow rolling' inflaton decay, if the
reheating time is of the order of a few Hubble time at the end
of inflation and if the reheating temperature depends only on at
what temperature the inflation sets in. Then, in these models
also, the lower bound on the density fluctuations is a
consequence of the holographic principle.

Often the inflaton decay, relevent in the models of new
inflation type, is modelled by that of a massive scalar field
interacting with other fields. In such models, the reheating
time (the reheating temperature) has no relation to the Hubble
time at the end (the temperature at the beginning) of
inflation. Typically, the holographic principle is automatically
satisfied in these models with no further consequences.

Perhaps, it is that such models of inflaton decay may be
specific possibilities only, while the generic possibilities
have the reheating time (the reheating temperature) of the order
of the Hubble time at the end (the temperature at the beginning)
of inflation. If so then, in the models of new inflation
type also, the holographic principle is likely to lead to a
lower bound on the denisty fluctuations.

Conversely, and just as likely, the inflaton decay models {\em
are} the generic models of reheating. Moreover, it may also be
that similar generic models exist for bubble wall collisions
also, in which the reheating time (the reheating temperature)
has no relation to the Hubble time at the end (the temperature
at the beginning) of inflation. If so then, in the models of
extended inflation type also, the holographic principle is
likely to be automatically satisfied with no further
consequences. It is desireable to settle this issue
definitively, but it is beyond the scope of the present work.

In closing, we mention an interesting application of the present
analysis: Note that by an appropriate coordinate transformation
\cite{book}, the inflating universe can be cast as a static de
Sitter one. One can then translate the present analysis and
compare the results with those obtained for some anti de Sitter
spaces in \cite{sw}. This might provide some insights into the
holographic principle in static universes.


\end{document}